\documentclass[twocolumn,pra,showpacs,superscriptaddress,floatfix]{revtex4}

\newcommand{\e}{{\mathrm e}}
\newcommand{\I}{{\mathrm i}}
\newcommand{\ham}{{\mathcal H}}

\begin{document}
\title{Practical Pulse Engineering: Gradient Ascent Without Matrix Exponentiation}
\author{Gaurav Bhole}
\author{Jonathan A. Jones}\email{jonathan.jones@physics.ox.ac.uk}
\affiliation{Centre for Quantum Computation, Clarendon Laboratory, University of Oxford, Parks Road, OX1 3PU, United Kingdom}
\date{\today}
\begin{abstract}
Since 2005 there has been a huge growth in the use of engineered control pulses to perform desired quantum operations in systems such as NMR quantum information processors. These approaches, which build on the original gradient ascent pulse engineering (GRAPE) algorithm, remain computationally intensive because of the need to calculate matrix exponentials for each time step in the control pulse.  Here we discuss how the propagators for each time step can be approximated using the Trotter--Suzuki formula, and a further speed up achieved by avoiding unnecessary operations.  The resulting procedure can give a substantial speed gain with negligible cost in propagator error, providing a more practical approach to pulse engineering.
\end{abstract}
\maketitle

Quantum information processors encode information in two-level quantum systems (qubits) and manipulate this through a series of elementary unitary transformations (quantum logic gates) \cite{Bennett2000,NCbook}.  Quantum control seeks to implement some target unitary propagator $U$ in a quantum system with background Hamiltonian $\ham_0$ by applying some time-dependent Hamiltonian $\ham_1(t)$. The resulting operator can be written as
\begin{equation}
V={\mathcal T}\left(\exp\left[-\I\int \ham_0+\ham_1(t)\,\mathrm{d}t\right]\right)
\end{equation}
where ${\mathcal T}$ is the Dyson time-ordering operator.  To make progress beyond this formal solution it is usually necessary to replace this continuously varying Hamiltonian by a piecewise constant form, so that
\begin{equation}
V=V_n V_{n-1}\dots V_1,\quad V_j=\exp\left[-\I(\ham_0+\ham_j)\delta t\right], \label{eq:Vj}
\end{equation}
and to write the time-varying portion of the Hamiltonian as a weighted sum of a set of $p$ distinct control fields
\begin{equation}
\ham_j=\sum_{k=1}^{p} a_j^k\ham^k.
\end{equation}
Any particular control pulse can then be described by the corresponding set of amplitudes, $a_{j}^k$, and the time step $\delta t$, here taken as fixed.  The quality of a control pulse can be measured by its fidelity with the desired operation $U$
\begin{equation}
\Phi=|\langle U|V\rangle|^2
\end{equation}
where the Hilbert–-Schmidt inner product is defined by $\langle U|V\rangle=\textrm{tr}(U^\dag V)$, possibly normalised by the dimension of the operators \cite{NCbook}.  The optimal control problem is then to find the set of amplitudes which maximises this fidelity, usually in the presence of practical constraints on the magnitudes of the amplitudes and the total length of the sequence.  This is computationally challenging as the dimension of the underlying Hilbert space rises exponentially with the number of qubits to be controlled, although this difficulty can in some cases be reduced by using subsystems to simplify the calculations \cite{Ryan2008}.  One recent approach \cite{Lu2017a} is to use subsystem methods to find approximate control pulses and then optimise these directly using the quantum system itself.  Whatever approach is adopted, it is important to perform any computations as efficiently as possible.

\section{Gradient Ascent}
Practical algorithms to maximize a function usually rely on the calculation of gradients, built from values of $\partial\Phi/\partial a_{j}^k$, but early work on optimal control was hampered by the belief that calculation of each of the $np$ distinct elements of the gradient vector would require the full evaluation of $V$, and thus a total of $n^2p$ sub-propagators.  The design of control pulses was a major topic within Nuclear Magnetic Resonance (NMR) \cite{EBWbook}, but most NMR sequences were paramaterized by a small number of variables to keep the size of the computation manageable.

A breakthrough in 2005 was the realisation that it is not necessary to recalculate these sub-propagators, which can instead be stored and reused.  The resulting algorithm, known as GRadient Ascent Pulse Engineering (GRAPE) \cite{Khaneja2005}, is now widely used, as are many variants, such as second order GRAPE \cite{DeFouquieres2011}, Newton--Raphson GRAPE \cite{Goodwin2016}, and GRadient Ascent in Functional Space (GRAFS) \cite{Lucarelli2016}.  However, although this algorithm reduces the number of sub-propagator calculations from $n^2p$ to $n$, the calculation of each of the $n$ sub-propagators still requires a matrix exponential, which must be re-evaluated at each round of the search algorithm.

A variety of algorithms exist for calculating numerical matrix exponentials \cite{Moler2003}, but the most widely used methods combine scaling and squaring with Pad\'{e} approximants \cite{Higham2005,Al-Mohy2009}. Whatever approach is used, this large number of matrix exponentials remains a key computational bottleneck, and reducing the number of such operations would greatly speed up the entire process. Similar issues are encountered in other optimal control algorithms such as Krotov \cite{Maximov2008}, Lyapunov \cite{Hou2014}, and Chopped Random Basis Optimization (CRAB) \cite{Caneva2011}.  As we shall show below, however, when control pulses are made up from a large number of small rotations then the matrix exponentials can be efficiently approximated with negligible error.

To make further progress we will initially specialise to the case of a homonuclear NMR system of $q$ distinct spin-$1/2$ nuclei, that is a homonuclear NMR implementation of $q$ qubits, with a Hilbert space of dimension $N=2^q$.  Our efficient algorithm is not confined to this case, but it is useful to begin with a concrete example.  The background Hamiltonian can then be written as
\begin{equation}
\ham_0=\sum_{r=1}^{q} \omega_r I^z_r+\sum_{s<r} \omega_{rs}\mathbf{I}_r\cdot\mathbf{I}_s
\end{equation}
where $I^z_r$ indicates $\frac12\sigma_z$ on the $r^{\rm th}$ spin \cite{EBWbook}, and we are working in natural units such that $\hbar=1$.  In the case of weak coupling, that is $|\omega_{rs}|\ll|\omega_r-\omega_s|$, this can be approximated by
\begin{equation}
\ham_0=\sum_r \omega_r I^z_r+\sum_{s<r} \omega_{rs}I^z_r I^z_s.
\end{equation}
The control Hamiltonian is provide by an RF oscillating magnetic field, with controllable amplitude and phase, but a fixed frequency close to resonance with the spins, which all have similar resonance frequencies in a homonuclear system.  Transforming into a rotating frame, and making the rotating wave approximation, the control Hamiltonian is now the sum of two terms, each of which is applied identically to all the spins,
\begin{equation}
\ham_j=x_j F^x +y_j F^y,
\end{equation}
where $F^x=\sum_r I^x_r$ is the total $x$-operator on all spins, and similarly for $F^y$.  The amplitudes of these control Hamiltonians are $x_j=\alpha_j\cos\phi_j$ and $y_j=\alpha_j\sin\phi_j$, where $\alpha_j$ is the magnitude of the applied field and $\phi_j$ is its phase. The background Hamiltonian $\ham_0$ is unaffected, except that the individual spin frequencies are replaced by their offsets from the rotating frame frequency \cite{EBWbook}.

It appears necessary to use a full matrix exponential to evaluate each sub-propagator $V_j$ in Eqn.~\ref{eq:Vj}, because $\ham_j$ is the sum of two non-commuting terms, neither of which commutes with $\ham_0$.  The matrix exponential is, of course, simple to evaluate in the eigenbasis of $\ham_0+\ham_j$, but finding this eigenbasis is as difficult as evaluating the matrix exponential directly. Progress can be made with an appropriate phase transformation \cite{Bhole2016}, into a frame where  $\ham_j$ is aligned with the $x$-axis.  This transformation is diagonal in the computational basis, and so easy to evaluate and perform.  In particular we write
\begin{equation}
\e^{-\I(\ham_0+\ham_j)\delta t}=\e^{-\I\phi_j F^z}\e^{-\I (\ham_0+\ham^x_j) \delta t}\e^{\I\phi_j F^z}
\end{equation}
where $\ham^x_j=\alpha_jF^x$. However the inner matrix exponential still involves the sum of two non-commuting terms, and so appears to require a computationally intensive full matrix exponential.

\section{Approximating unitaries}
It is, however, possible to approximate the sub-propagator as long as $\delta t$ is small enough, as all small-angle unitary evolutions approximately commute.  If the terms are very small then it suffices to write
\begin{equation}
\e^{-\I(\ham_0+\ham^x_j)\delta t}\approx\e^{-\I\ham_0\delta t}\e^{-\I\ham^x_j\delta t}\approx\e^{-\I\ham^x_j\delta t}\e^{-\I\ham_0\delta t},
\end{equation}
but this Trotter approximation, implicitly used by Bhole and Mahesh \cite{Bhole2017}, is only accurate to second order in $\delta t$.  It is better to use the Trotter--Suzuki form \cite{NCbook,Suzuki1986}
\begin{equation}
\e^{-\I(\ham_0+\ham^x_j)\delta t}\approx\e^{-\I\ham_0\delta t/2}\e^{-\I\ham^x_j\delta t}\e^{-\I\ham_0\delta t/2},
\end{equation}
which is accurate to third order in $\delta t$.  As the propagator fidelity depends quadratically on the size of the error, this approximation will be accurate to sixth order in $\delta t$ when calculating fidelities and their associated gradients.  A more thorough treatment of these errors is given below, but in our experience these errors are essentially negligible for pulse engineering calculations.

It might seem that approximating a matrix exponential by the product of three exponentials does not constitute progress.  However, the two outer terms are constant, and so only need to be evaluated once for the entire GRAPE calculation, while the central Hamiltonian is a simple scalar multiple of $F^x$, and so will always be diagonal in the Hadamard basis. The basis transformation can be combined with the fixed outer terms, giving
\begin{equation}
V_j\approx\e^{-\I\phi_j F^z}W_1\e^{-\I \alpha_j F^z \delta t}W_2\e^{\I\phi_j F^z}, \label{eq:approxVj}
\end{equation}
with all explicit matrix exponentials now diagonal in the computational basis. The basis transformations
\begin{equation}
W_1=\e^{-\I\ham_0\delta t/2}{\rm H}^{(q)},\quad W_2={\rm H}^{(q)}\e^{-\I\ham_0\delta t/2}, \label{eq:bases}
\end{equation}
where ${\rm H}^{(q)}$ is the $q$-qubit Hadamard gate, are the same for every propagator and so are evaluated only once, requiring only a single full matrix exponential.

This expression does not depend on the weak coupling approximation, as strong couplings are unaffected by phase transformations. Similarly dipolar couplings are axially symmetric, and so once again unaffected. It can be trivially extended to heteronuclear spin systems, as control fields applied to one nuclear species do not affect spins of other species, and so all commute with one another.  The resulting approximate expression is \textit{identical} to Eqn.~\ref{eq:approxVj}, except that the three diagonal matrix exponentials are now calculated over weighted sums of operators corresponding to each control field.

\section{Errors and speed gains}
The errors in this approach are most simply considered for a one-spin system with $\ham_0=\omega_0 I^z$ and $\ham_j=\alpha_j I^x$, which permits analytical calculations.  The fidelity between the exact value of $V_j$, Eqn.~\ref{eq:Vj}, and its approximation, Eqn.~\ref{eq:approxVj}, is
\begin{equation}
\Phi=1-\frac{\omega_0^2\alpha_j^2(\omega_0^2+4\alpha_j^2)}{2304}\delta t^6+O(\delta t^8). \label{eq:onespin}
\end{equation}
For realistic frequencies and time steps in NMR systems, the infidelity of the approximate approach will be very small.  Offset frequencies rarely exceed 15\,kHz, and for the low RF powers used during long control pulses the nutation rate is usually below 5\,kHz.  Even for these extreme values, the error for a $10\,\mu\rm s$ time step is below $5\times10^{-5}$, and for the lower frequencies and smaller time steps normally used the error will far smaller.

Evaluating the error for pulse engineering in a real system is more complicated, reflecting the larger matrices involved, the large number of relevant frequencies, and the need to evaluate the error in the entire combined propagator, and not just the individual sub-propagators.  The worst case occurs when each sub-propagator is identical: in this case the errors grow linearly with the number of steps, and so the infidelity, which depends on the square of the error, grows quadratically with the number of steps. However such a case is quite unrealistic in practice, as variations in the amplitude and phase of control fields mean that errors will partly cancel. Our simulations indicate that the infidelities remain small in realistic cases, typically around $10^{-4}$.

The speed gains achieved by this approximate approach arise from avoiding full matrix exponentials by performing all calculations in an appropriate eigenbasis where the operators are diagonal.  The new computational bottleneck is the calculation of matrix products,  and since both matrix multiplication and numerical matrix exponentiation have a computational complexity of $O(N^3)$ for $N$-by-$N$ matrices, these gains are approximately constant, and independent of the dimension of the Hamiltonian. (Much larger speed gains have been demonstrated in open system GRAPE \cite{Boutin2017}, but only for state-to-state tasks.)

As three of the matrices in Eqn.~\ref{eq:approxVj} are diagonal most steps can be carried out using efficient partial matrix multiplication, and only one full matrix product is required (see the Appendix).  The exact speed up achieved can be quite complex, due to implicit parallel computation on larger matrices.  While this reduces the wall clock time, the CPU time will be increased to handle the overheads of parallelisation. This will affect the observed speed-up in a way that depends upon matrix size \cite{Waldherr2010} as matrix exponentiation is more easily parallelised than our efficient algorithm. The speed gain achievable will depend on both the spin system chosen and the precise code used, but our MATLAB simulations for the 3-spin homonuclear system corresponding to the three $^{13}\textrm{C}$ nuclei in alanine \cite{Cory1998} indicate that the calculation of sub-propagators can be sped up by a factor of around 18, while full propagators (which require an additional full matrix multiplication for each sub-propagator) are sped up by a factor of around 13.  This speed gain means that a four-spin system can be simulated more rapidly with our approximate approach than an exact simulation of a three-spin system.

\section{Reducing errors}
The infidelity in Eqn.~\ref{eq:approxVj} arises from the fact that $\ham^x_j$ does not commute with $\ham_0$, and can be minimised by making $\ham^x_j$ as small as possible.  This can be achieved by moving part of $\ham^x_j$ into $\ham_0$, writing
\begin{equation}
\ham_0'=\ham_0+\Omega F^x, \quad \ham_j'=\alpha_j' F^x = (\alpha_j-\Omega) F^x,
\end{equation}
where the offset frequency $\Omega$ is chosen to make the values $\alpha_j'$ as close to zero as possible, for example by setting it to half the maximum amplitude expected. While this will increase the infidelity of some individual sub-propagators, 
on average the infidelity will be decreased, and the fidelity of the total propagator will improve.  The basis transformation operators, Eqn.~\ref{eq:bases}, have to be calculated for the new value of $\ham_0'$, but as this only needs to be done once for the entire calculation this gain in fidelity comes at no cost in time.

A more accurate approach is to choose $\Omega$ as the mean amplitude for the particular propagator being calculated: in this case the basis transformation operators have to be recalculated each time, but this still only requires one full matrix exponential for each propagator.  If even higher accuracy is required then it is possible to use two different offset values, one for small values of $\alpha_j$ and one for large values, choosing the appropriate basis transformation in each case.  Our simulations suggest that using a single value of $\Omega$ can improve the infidelity of the propagator by a factor around 15, while using two values can give an improvement of about 200, leading to propagator infidelities around $10^{-6}$.

This could be improved still further by using a larger number of offsets, effectively trading memory for time \cite{Bhole2017}, but these later gains are smaller than the early ones.  When the numbers of offsets is large then all values of $\alpha'_j$ are small in comparison with all other frequencies, and in this limit the infidelity falls quadratically with the number of offsets, as expected from Eqn.~\ref{eq:onespin}.  If very precise pulse engineering is required then it is simpler to use approximate techniques in the early stages of optimization, and switch to exact calculations using full matrix exponentials once the pulse fidelity is high enough to justify this.

\section{Conclusions}
We have applied our efficient algorithm with a single offset frequency to calculate GRAPE control pulses in a variety of NMR systems, observing a speed increase of at least a factor of 6.  We have checked each pulse against a conventional full matrix exponential calculation, and the error in the fidelity of the final propagator has always been below $10^{-4}$.  As typical experiments in this field seek a fidelity of around 0.999 \cite{Xin2018a} this error is effectively negligible, and approximate propagators provide an entirely practical approach to pulse engineering.

Although developed, like GRAPE, for use in NMR, our approach could be used in other fields where GRAPE has been applied such as ESR \cite{Zhang2011a}, NV centres \cite{Dolde2014}, ion traps \cite{Nebendahl2009}, and circuit QED \cite{Fisher2010}.  The key requirement is that control Hamiltonians either commute with one another or can be converted to commuting forms using phase transformations which commute with the background Hamiltonian, and that the individual sub-propagators remain small enough to use Trotter--Suzuki approximations.

\begin{acknowledgments}
G.~Bhole is supported by a Felix Scholarship.
\end{acknowledgments}

\appendix*
\section{Efficient matrix multiplication}
As matrix multiplication is now the computational bottleneck, it is vital that this is carried out as efficiently as possible.  Full matrix multiplication, with computational complexity of $O(N^3)$, should only be used when actually necessary: when multiplying a full matrix by a diagonal matrix only $O(N^2)$ steps are required.  

If the algorithm is coded in a low level language then it is easy to ensure this, but in higher level languages, such as MATLAB, it is necessary to code carefully.  Diagonal matrices should be stored not as matrices but as vectors, to avoid unnecessary operations.  In particular the exponentials of diagonal matrices must be calculated using direct exponentiation of the individual elements.  To combine the individual matrices,  Eqn.~\ref{eq:approxVj} can be written as
\begin{equation}
\{\phi\}W_1\{\alpha\}W_2\{\phi^*\}
\end{equation}
where braces indicate diagonal matrices stored as vectors.  The multiplications by the two outer diagonal matrices can be combined, and the overall process written as
\begin{equation}
\Phi\,.\!*(W_1*(\alpha\,.\!*W_2)), \quad\textrm{with}\quad \Phi=\phi\,.\!*\phi',
\end{equation}
where $*$ is the MATLAB operator for a full matrix multiplication, $.*$ is the operator for an element-by-element multiplication, and $\phi'$ indicates the adjoint of the vector $\phi$.  Note that only a single full matrix multiplication is required.

Combining the phase multiplications into a single matrix $\Phi$ is particularly useful when developing pulses which are robust to variations in the RF coupling strength.  These are simulated by evaluating propagators with the control amplitude set to a range of values \cite{Xin2018a} (e.g., 95\%, 100\% and 105\% of the nominal value).  In such cases the matrix $\Phi$ is the same for all the different coupling strengths, and need only be calculated once.

When programming in MATLAB it is also important to think carefully about memory handling. In particular it is quicker to evaluate Eqn.~A.2 one multiplication at a time, storing intermediate results in explicit variables.  If the multiplications are carried out in one line then temporary variables are created to hold intermediate results, and subsequently destroyed.  If equivalent calculations are carried out many times, as happens when evaluating propagators, it is quicker to reuse previously allocated variables.  These minor issues are almost irrelevant in conventional GRAPE calculations, where the time needed for matrix exponentiation dominates over everything else, but become important once all these slow stages have been removed.

It might appear possible to speed up calculations further, for example by using the structure in $F^z$ to avoid repeatedly calculating the same exponential terms.  This would certainly be sensible when programming in a low level language, but in our experience such tricks actually slow MATLAB down.  It can be difficult to predict precisely what will give the fastest MATLAB code, and experimentation is the best approach.

\bibliography{GRAWME}

\begin{thebibliography}{26}
\expandafter\ifx\csname natexlab\endcsname\relax\def\natexlab#1{#1}\fi
\expandafter\ifx\csname bibnamefont\endcsname\relax
  \def\bibnamefont#1{#1}\fi
\expandafter\ifx\csname bibfnamefont\endcsname\relax
  \def\bibfnamefont#1{#1}\fi
\expandafter\ifx\csname citenamefont\endcsname\relax
  \def\citenamefont#1{#1}\fi
\expandafter\ifx\csname url\endcsname\relax
  \def\url#1{\texttt{#1}}\fi
\expandafter\ifx\csname urlprefix\endcsname\relax\def\urlprefix{URL }\fi
\providecommand{\bibinfo}[2]{#2}
\providecommand{\eprint}[2][]{\url{#2}}

\bibitem[{\citenamefont{Bennett and DiVincenzo}(2000)}]{Bennett2000}
\bibinfo{author}{\bibfnamefont{C.~H.} \bibnamefont{Bennett}} \bibnamefont{and}
  \bibinfo{author}{\bibfnamefont{D.~P.} \bibnamefont{DiVincenzo}},
  \bibinfo{journal}{Nature} \textbf{\bibinfo{volume}{404}},
  \bibinfo{pages}{247} (\bibinfo{year}{2000}).

\bibitem[{\citenamefont{Nielsen and Chuang}(2000)}]{NCbook}
\bibinfo{author}{\bibfnamefont{M.~A.} \bibnamefont{Nielsen}} \bibnamefont{and}
  \bibinfo{author}{\bibfnamefont{I.~L.} \bibnamefont{Chuang}},
  \emph{\bibinfo{title}{Quantum Computation and Quantum Information}}
  (\bibinfo{publisher}{CUP}, \bibinfo{year}{2000}).

\bibitem[{\citenamefont{Ryan et~al.}(2008)\citenamefont{Ryan, Negrevergne,
  Laforest, Knill, and Laflamme}}]{Ryan2008}
\bibinfo{author}{\bibfnamefont{C.~A.} \bibnamefont{Ryan}},
  \bibinfo{author}{\bibfnamefont{C.}~\bibnamefont{Negrevergne}},
  \bibinfo{author}{\bibfnamefont{M.}~\bibnamefont{Laforest}},
  \bibinfo{author}{\bibfnamefont{E.}~\bibnamefont{Knill}}, \bibnamefont{and}
  \bibinfo{author}{\bibfnamefont{R.}~\bibnamefont{Laflamme}},
  \bibinfo{journal}{Phys. Rev. A} \textbf{\bibinfo{volume}{78}},
  \bibinfo{pages}{012328} (\bibinfo{year}{2008}).

\bibitem[{\citenamefont{Lu et~al.}(2017)\citenamefont{Lu, Li, Li, Katiyar,
  Park, Feng, Xin, Li, Long, Brodutch et~al.}}]{Lu2017a}
\bibinfo{author}{\bibfnamefont{D.}~\bibnamefont{Lu}},
  \bibinfo{author}{\bibfnamefont{K.}~\bibnamefont{Li}},
  \bibinfo{author}{\bibfnamefont{J.}~\bibnamefont{Li}},
  \bibinfo{author}{\bibfnamefont{H.}~\bibnamefont{Katiyar}},
  \bibinfo{author}{\bibfnamefont{A.~J.} \bibnamefont{Park}},
  \bibinfo{author}{\bibfnamefont{G.}~\bibnamefont{Feng}},
  \bibinfo{author}{\bibfnamefont{T.}~\bibnamefont{Xin}},
  \bibinfo{author}{\bibfnamefont{H.}~\bibnamefont{Li}},
  \bibinfo{author}{\bibfnamefont{G.}~\bibnamefont{Long}},
  \bibinfo{author}{\bibfnamefont{A.}~\bibnamefont{Brodutch}},
  \bibnamefont{et~al.}, \bibinfo{journal}{npj Quantum Information}
  \textbf{\bibinfo{volume}{3}}, \bibinfo{pages}{45} (\bibinfo{year}{2017}).

\bibitem[{\citenamefont{Ernst et~al.}(1987)\citenamefont{Ernst, Bodenhausen,
  and Wokaun}}]{EBWbook}
\bibinfo{author}{\bibfnamefont{R.~R.} \bibnamefont{Ernst}},
  \bibinfo{author}{\bibfnamefont{G.}~\bibnamefont{Bodenhausen}},
  \bibnamefont{and} \bibinfo{author}{\bibfnamefont{A.}~\bibnamefont{Wokaun}},
  \emph{\bibinfo{title}{Principles of Nuclear Magnetic Resonance in One and Two
  Dimensions}} (\bibinfo{publisher}{Oxford University Press},
  \bibinfo{year}{1987}).

\bibitem[{\citenamefont{Khaneja et~al.}(2005)\citenamefont{Khaneja, Reiss,
  Kehlet, Schulte-Herbr{\"u}ggen, and Glaser}}]{Khaneja2005}
\bibinfo{author}{\bibfnamefont{N.}~\bibnamefont{Khaneja}},
  \bibinfo{author}{\bibfnamefont{T.}~\bibnamefont{Reiss}},
  \bibinfo{author}{\bibfnamefont{C.}~\bibnamefont{Kehlet}},
  \bibinfo{author}{\bibfnamefont{T.}~\bibnamefont{Schulte-Herbr{\"u}ggen}},
  \bibnamefont{and} \bibinfo{author}{\bibfnamefont{S.~J.}
  \bibnamefont{Glaser}}, \bibinfo{journal}{J. Magn. Reson.}
  \textbf{\bibinfo{volume}{172}}, \bibinfo{pages}{296} (\bibinfo{year}{2005}).

\bibitem[{\citenamefont{De~Fouquieres et~al.}(2011)\citenamefont{De~Fouquieres,
  Schirmer, Glaser, and Kuprov}}]{DeFouquieres2011}
\bibinfo{author}{\bibfnamefont{P.}~\bibnamefont{De~Fouquieres}},
  \bibinfo{author}{\bibfnamefont{S.~G.} \bibnamefont{Schirmer}},
  \bibinfo{author}{\bibfnamefont{S.~J.} \bibnamefont{Glaser}},
  \bibnamefont{and} \bibinfo{author}{\bibfnamefont{I.}~\bibnamefont{Kuprov}},
  \bibinfo{journal}{J. Magn. Reson.} \textbf{\bibinfo{volume}{212}},
  \bibinfo{pages}{412} (\bibinfo{year}{2011}).

\bibitem[{\citenamefont{Goodwin and Kuprov}(2016)}]{Goodwin2016}
\bibinfo{author}{\bibfnamefont{D.~L.} \bibnamefont{Goodwin}} \bibnamefont{and}
  \bibinfo{author}{\bibfnamefont{I.}~\bibnamefont{Kuprov}},
  \bibinfo{journal}{J. Chem. Phys.} \textbf{\bibinfo{volume}{144}},
  \bibinfo{pages}{204107} (\bibinfo{year}{2016}).

\bibitem[{\citenamefont{Lucarelli}(2016)}]{Lucarelli2016}
\bibinfo{author}{\bibfnamefont{D.~G.} \bibnamefont{Lucarelli}},
  \bibinfo{journal}{preprint arXiv:1611.00188}  (\bibinfo{year}{2016}).

\bibitem[{\citenamefont{Moler and Loan}(2003)}]{Moler2003}
\bibinfo{author}{\bibfnamefont{C.}~\bibnamefont{Moler}} \bibnamefont{and}
  \bibinfo{author}{\bibfnamefont{C.~V.} \bibnamefont{Loan}},
  \bibinfo{journal}{SIAM Review} \textbf{\bibinfo{volume}{45}},
  \bibinfo{pages}{3} (\bibinfo{year}{2003}).

\bibitem[{\citenamefont{Higham}(2005)}]{Higham2005}
\bibinfo{author}{\bibfnamefont{N.~J.} \bibnamefont{Higham}},
  \bibinfo{journal}{SIAM J. Matrix Anal. Appl.} \textbf{\bibinfo{volume}{26}},
  \bibinfo{pages}{1179} (\bibinfo{year}{2005}).

\bibitem[{\citenamefont{Al-Mohy and Higham}(2009)}]{Al-Mohy2009}
\bibinfo{author}{\bibfnamefont{A.~H.} \bibnamefont{Al-Mohy}} \bibnamefont{and}
  \bibinfo{author}{\bibfnamefont{N.~J.} \bibnamefont{Higham}},
  \bibinfo{journal}{SIAM J. Matrix Anal. Appl.} \textbf{\bibinfo{volume}{31}},
  \bibinfo{pages}{970} (\bibinfo{year}{2009}).

\bibitem[{\citenamefont{Maximov et~al.}(2008)\citenamefont{Maximov,
  To{\v{s}}ner, and Nielsen}}]{Maximov2008}
\bibinfo{author}{\bibfnamefont{I.~I.} \bibnamefont{Maximov}},
  \bibinfo{author}{\bibfnamefont{Z.}~\bibnamefont{To{\v{s}}ner}},
  \bibnamefont{and} \bibinfo{author}{\bibfnamefont{N.~C.}
  \bibnamefont{Nielsen}}, \bibinfo{journal}{J. Chem. Phys.}
  \textbf{\bibinfo{volume}{128}}, \bibinfo{pages}{05B609}
  (\bibinfo{year}{2008}).

\bibitem[{\citenamefont{Hou et~al.}(2014)\citenamefont{Hou, Wang, and
  Yi}}]{Hou2014}
\bibinfo{author}{\bibfnamefont{S.~C.} \bibnamefont{Hou}},
  \bibinfo{author}{\bibfnamefont{L.~C.} \bibnamefont{Wang}}, \bibnamefont{and}
  \bibinfo{author}{\bibfnamefont{X.~X.} \bibnamefont{Yi}},
  \bibinfo{journal}{Phys. Lett. A} \textbf{\bibinfo{volume}{378}},
  \bibinfo{pages}{699 } (\bibinfo{year}{2014}).

\bibitem[{\citenamefont{Caneva et~al.}(2011)\citenamefont{Caneva, Calarco, and
  Montangero}}]{Caneva2011}
\bibinfo{author}{\bibfnamefont{T.}~\bibnamefont{Caneva}},
  \bibinfo{author}{\bibfnamefont{T.}~\bibnamefont{Calarco}}, \bibnamefont{and}
  \bibinfo{author}{\bibfnamefont{S.}~\bibnamefont{Montangero}},
  \bibinfo{journal}{Phys. Rev. A} \textbf{\bibinfo{volume}{84}},
  \bibinfo{pages}{022326} (\bibinfo{year}{2011}).

\bibitem[{\citenamefont{Bhole et~al.}(2016)\citenamefont{Bhole, Anjusha, and
  Mahesh}}]{Bhole2016}
\bibinfo{author}{\bibfnamefont{G.}~\bibnamefont{Bhole}},
  \bibinfo{author}{\bibfnamefont{V.~S.} \bibnamefont{Anjusha}},
  \bibnamefont{and} \bibinfo{author}{\bibfnamefont{T.~S.}
  \bibnamefont{Mahesh}}, \bibinfo{journal}{Phys. Rev. A}
  \textbf{\bibinfo{volume}{93}}, \bibinfo{pages}{042339}
  (\bibinfo{year}{2016}).

\bibitem[{\citenamefont{Bhole and Mahesh}(2017)}]{Bhole2017}
\bibinfo{author}{\bibfnamefont{G.}~\bibnamefont{Bhole}} \bibnamefont{and}
  \bibinfo{author}{\bibfnamefont{T.~S.} \bibnamefont{Mahesh}},
  \bibinfo{journal}{preprint arXiv:1707.02162}  (\bibinfo{year}{2017}).

\bibitem[{\citenamefont{Suzuki}(1986)}]{Suzuki1986}
\bibinfo{author}{\bibfnamefont{M.}~\bibnamefont{Suzuki}}, \bibinfo{journal}{J.
  Stat. Phys.} \textbf{\bibinfo{volume}{43}}, \bibinfo{pages}{883}
  (\bibinfo{year}{1986}).

\bibitem[{\citenamefont{Waldherr et~al.}(2010)\citenamefont{Waldherr, Huckle,
  Auckenthaler, Sander, and Schulte-Herbr{\"u}ggen}}]{Waldherr2010}
\bibinfo{author}{\bibfnamefont{K.}~\bibnamefont{Waldherr}},
  \bibinfo{author}{\bibfnamefont{T.}~\bibnamefont{Huckle}},
  \bibinfo{author}{\bibfnamefont{T.}~\bibnamefont{Auckenthaler}},
  \bibinfo{author}{\bibfnamefont{U.}~\bibnamefont{Sander}}, \bibnamefont{and}
  \bibinfo{author}{\bibfnamefont{T.}~\bibnamefont{Schulte-Herbr{\"u}ggen}},
  \emph{\bibinfo{title}{High Performance Computing in Science and Engineering}}
  (\bibinfo{publisher}{Springer}, \bibinfo{year}{2010}), chap.
  \bibinfo{chapter}{Fast 3D Block Parallelisation for the Matrix Multiplication
  Prefix Problem}, pp. \bibinfo{pages}{39--50}.

\bibitem[{\citenamefont{Cory et~al.}(1998)\citenamefont{Cory, Price, Maas,
  Knill, Laflamme, Zurek, Havel, and Somaroo}}]{Cory1998}
\bibinfo{author}{\bibfnamefont{D.~G.} \bibnamefont{Cory}},
  \bibinfo{author}{\bibfnamefont{M.~D.} \bibnamefont{Price}},
  \bibinfo{author}{\bibfnamefont{W.}~\bibnamefont{Maas}},
  \bibinfo{author}{\bibfnamefont{E.}~\bibnamefont{Knill}},
  \bibinfo{author}{\bibfnamefont{R.}~\bibnamefont{Laflamme}},
  \bibinfo{author}{\bibfnamefont{W.~H.} \bibnamefont{Zurek}},
  \bibinfo{author}{\bibfnamefont{T.~F.} \bibnamefont{Havel}}, \bibnamefont{and}
  \bibinfo{author}{\bibfnamefont{S.~S.} \bibnamefont{Somaroo}},
  \bibinfo{journal}{Phys. Rev. Lett.} \textbf{\bibinfo{volume}{81}},
  \bibinfo{pages}{2152} (\bibinfo{year}{1998}).

\bibitem[{\citenamefont{Boutin et~al.}(2017)\citenamefont{Boutin, Andersen,
  Venkatraman, Ferris, and Blais}}]{Boutin2017}
\bibinfo{author}{\bibfnamefont{S.}~\bibnamefont{Boutin}},
  \bibinfo{author}{\bibfnamefont{C.~K.} \bibnamefont{Andersen}},
  \bibinfo{author}{\bibfnamefont{J.}~\bibnamefont{Venkatraman}},
  \bibinfo{author}{\bibfnamefont{A.~J.} \bibnamefont{Ferris}},
  \bibnamefont{and} \bibinfo{author}{\bibfnamefont{A.}~\bibnamefont{Blais}},
  \bibinfo{journal}{Phys. Rev. A} \textbf{\bibinfo{volume}{96}},
  \bibinfo{pages}{042315} (\bibinfo{year}{2017}).

\bibitem[{\citenamefont{Xin et~al.}(2018)\citenamefont{Xin, Huang, Lu, Li, Luo,
  Yin, Li, Lu, Long, and Zeng}}]{Xin2018a}
\bibinfo{author}{\bibfnamefont{T.}~\bibnamefont{Xin}},
  \bibinfo{author}{\bibfnamefont{S.}~\bibnamefont{Huang}},
  \bibinfo{author}{\bibfnamefont{S.}~\bibnamefont{Lu}},
  \bibinfo{author}{\bibfnamefont{K.}~\bibnamefont{Li}},
  \bibinfo{author}{\bibfnamefont{Z.}~\bibnamefont{Luo}},
  \bibinfo{author}{\bibfnamefont{Z.}~\bibnamefont{Yin}},
  \bibinfo{author}{\bibfnamefont{J.}~\bibnamefont{Li}},
  \bibinfo{author}{\bibfnamefont{D.}~\bibnamefont{Lu}},
  \bibinfo{author}{\bibfnamefont{G.}~\bibnamefont{Long}}, \bibnamefont{and}
  \bibinfo{author}{\bibfnamefont{B.}~\bibnamefont{Zeng}},
  \bibinfo{journal}{Sci. Bull.} \textbf{\bibinfo{volume}{63}},
  \bibinfo{pages}{17 } (\bibinfo{year}{2018}).

\bibitem[{\citenamefont{Zhang et~al.}(2011)\citenamefont{Zhang, Ryan, Laflamme,
  and Baugh}}]{Zhang2011a}
\bibinfo{author}{\bibfnamefont{Y.}~\bibnamefont{Zhang}},
  \bibinfo{author}{\bibfnamefont{C.~A.} \bibnamefont{Ryan}},
  \bibinfo{author}{\bibfnamefont{R.}~\bibnamefont{Laflamme}}, \bibnamefont{and}
  \bibinfo{author}{\bibfnamefont{J.}~\bibnamefont{Baugh}},
  \bibinfo{journal}{Phys. Rev. Lett.} \textbf{\bibinfo{volume}{107}},
  \bibinfo{pages}{170503} (\bibinfo{year}{2011}).

\bibitem[{\citenamefont{Dolde et~al.}(2014)\citenamefont{Dolde, Bergholm, Wang,
  Jakobi, Naydenov, Pezzagna, Meijer, Jelezko, Neumann, Schulte-Herbrüggen
  et~al.}}]{Dolde2014}
\bibinfo{author}{\bibfnamefont{F.}~\bibnamefont{Dolde}},
  \bibinfo{author}{\bibfnamefont{V.}~\bibnamefont{Bergholm}},
  \bibinfo{author}{\bibfnamefont{Y.}~\bibnamefont{Wang}},
  \bibinfo{author}{\bibfnamefont{I.}~\bibnamefont{Jakobi}},
  \bibinfo{author}{\bibfnamefont{B.}~\bibnamefont{Naydenov}},
  \bibinfo{author}{\bibfnamefont{S.}~\bibnamefont{Pezzagna}},
  \bibinfo{author}{\bibfnamefont{J.}~\bibnamefont{Meijer}},
  \bibinfo{author}{\bibfnamefont{F.}~\bibnamefont{Jelezko}},
  \bibinfo{author}{\bibfnamefont{P.}~\bibnamefont{Neumann}},
  \bibinfo{author}{\bibfnamefont{T.}~\bibnamefont{Schulte-Herbrüggen}},
  \bibnamefont{et~al.}, \bibinfo{journal}{Nat. Commun.}
  \textbf{\bibinfo{volume}{5}}, \bibinfo{pages}{3371} (\bibinfo{year}{2014}).

\bibitem[{\citenamefont{Nebendahl et~al.}(2009)\citenamefont{Nebendahl,
  H\"affner, and Roos}}]{Nebendahl2009}
\bibinfo{author}{\bibfnamefont{V.}~\bibnamefont{Nebendahl}},
  \bibinfo{author}{\bibfnamefont{H.}~\bibnamefont{H\"affner}},
  \bibnamefont{and} \bibinfo{author}{\bibfnamefont{C.~F.} \bibnamefont{Roos}},
  \bibinfo{journal}{Phys. Rev. A} \textbf{\bibinfo{volume}{79}},
  \bibinfo{pages}{012312} (\bibinfo{year}{2009}).

\bibitem[{\citenamefont{Fisher et~al.}(2010)\citenamefont{Fisher, Helmer,
  Glaser, Marquardt, and Schulte-Herbr\"uggen}}]{Fisher2010}
\bibinfo{author}{\bibfnamefont{R.}~\bibnamefont{Fisher}},
  \bibinfo{author}{\bibfnamefont{F.}~\bibnamefont{Helmer}},
  \bibinfo{author}{\bibfnamefont{S.~J.} \bibnamefont{Glaser}},
  \bibinfo{author}{\bibfnamefont{F.}~\bibnamefont{Marquardt}},
  \bibnamefont{and}
  \bibinfo{author}{\bibfnamefont{T.}~\bibnamefont{Schulte-Herbr\"uggen}},
  \bibinfo{journal}{Phys. Rev. B} \textbf{\bibinfo{volume}{81}},
  \bibinfo{pages}{085328} (\bibinfo{year}{2010}).

\end{thebibliography}

\end{document}